\documentclass[ superscriptaddress, twocolumn, secnumarabic, tightenlines, nobibnotes, aps, prl]{revtex4}
\usepackage{bm}
\usepackage{amsmath,amssymb,amsfonts,epsfig,graphicx}

\usepackage{colordvi}
\begin{document}

\title{High-temperature series expansion of the Helmholtz free energy 
of the quantum spin-S XYZ chain}

\date{\today}

\author{Onofre Rojas}
\affiliation{Departamento de Ci\^encias Exatas, Universidade Federal de Lavras, Caixa Postal 37, 
    CEP 37200-000, Lavras-MG,  Brazil}
\author{S.M. de Souza}
\affiliation{Departamento de Ci\^encias Exatas, Universidade Federal de Lavras, Caixa Postal 37, 
CEP 37200-000, Lavras-MG,  Brazil}
\author{E.V. Corr\^ea Silva}
\affiliation{Departamento de Matem\'atica e Computa\c c\~ao, Faculdade de Tecnologia, 
Universidade do Estado do Rio de Janeiro. Estrada Resende-Riachuelo s/n$^{\textit o}$, 
Morada da Colina, CEP 27523-000,  Resende-RJ, Brazil}
\author{M.T. Thomaz}
\affiliation{Instituto de F\'{\i}sica, Universidade Federal Fluminense, Av. Gal. Milton Tavares de 
Souza s/n$^{\textit o}$, CEP 24210-340, Niter\'oi - RJ, Brazil}

\begin{abstract}
We consider the $XYZ$ chain model of arbitrary spin $S$ in the high
temperature region, with external magnetic field and single-ion
anisotropy term. Our high-temperature expansion of the Helmholtz free
energy is analytic in the parameters of the model for $S$, which may
range from $1/2$ to the classical limit of infinite spin ($S\rightarrow \infty$). Our
expansion is carried out up to order $(J\,\beta)^5$. Our results agree with
numerical results of the specific heat per site 
for $S=1/2$, obtained by the Bethe ansatz,  with $h=0$
and $D=0$. Finally, we show that the magnetic susceptibility and
magnetization of the quantum model can be well approximated 
by their classical analog in this region of temperature.

\end{abstract}

\maketitle

The one-dimensional Heisenberg model describing the exchange interaction of spins
is intensively applied to study the magnetic properties of 
quantum spin chains.
The XXZ spin-1/2 chain with external magnetic field along the $z$
direction is exactly integrable, consequently 
its thermodynamics has been investigated since very early works 
\cite{takahashi70, klumper93}. It has also been verified the
existence of higher-spin quasi one-dimensional magnetic systems such
as ${\rm CsVCl}_3$ and ${\rm CsVBr}_3$ ($S=3/2$)\cite{kadowaki,itoh},
$({\rm C}_{10}{\rm H}_8{\rm N}_2){\rm MnCl}_3$ ($S=2$)\cite{granroth}
and $({\rm CD}_3)_4{\rm NMnCl}_3$ $(S=5/2)$\cite{birgeneau,hutchings}.
All these antiferromagnetic structures exhibit nearly ideal
one-dimensional behavior over a considerable range of temperature.
Motivated by these features, magnetic and thermodynamical properties
for higher-spin chains were first investigated
numerically\cite{n_spins}. More recently we obtained in Ref.
\cite{ejpB} the $\beta$-expansion of the Helmholtz 
free energy (HFE) of the $XXZ$ model, 
for arbitrary spin-$S$, up to order ($J\, \beta)^6$, where $\beta = 1/kT$
($k$ is the Boltzmann constant and $T$ is the absolute temperature), 
in the presence of an external magnetic field and a single-ion anisotropy  term. 
We showed that in the high temperature region the magnetization and the
magnetic susceptibility, even for low values of spin 
such as $S=3/2$, can be
properly approximated by the classical results within 
a percental error of less than $4\%$.

Quasi-one-dimensional magnetic systems have been synthesized
experimentally such 
as the $({\rm C_6 H_{11}NH_3)CuBr_3}$ 
and the $({\rm C_6 H_{11}NH_3)CuCl_3}$, known as CHAB and
CHAC, respectively\cite{phaff84},
both exhibiting small anisotropy in the easy plane.
They are described by the spin-$1/2$  XYZ spin model. 
By the appropriate design of molecules, it is possible to 
obtain a variety of spin systems, including 
those with spin larger than $1/2$.

The thermodynamics of the $XYZ$ model has been solved
exactly for the spin-$1/2$ case without external magnetic fields; 
in this case, the model is integrable and its HFE is obtained
from the Bethe ansatz 
through a couple of integral equations\cite{klumper93}
and an infinite non-linear coupled equation\cite{takahashi91}.

The model is no longer integrable, though, if an external magnetic
field in any direction is introduced. This is the case of the XXZ
spin-1/2 spin chain with transverse external magnetic field; so far,
only its phase diagram at zero temperature has been
investigated\cite{dmitriev}, since the external magnetic field does
not commute with the Hamiltonian. However, an analytic high
temperature series of its HFE has not been obtained so far.

In this communication we present  the high temperature thermodynamics
of the  spin-$S$ $XYZ$ model, for arbitrary value of $S$, in the presence 
of an external magnetic field and a 
single-ion anisotropy term. We also investigate
if the anisotropy in the $XY$  plane avoids or 
does not avoid the classical behavior of the
magnetic susceptibility and magnetization in this region of temperature, even
for $S=3/2$, as was  shown in Ref.  \cite{ejpB} for the spin-$S$ XXZ model.

The Hamiltonian  of the spin-$S$ XYZ quantum 
periodic chain with $N$ sites  reads  

\begin{align}
{\bf H} =& \sum_{i=1}^N \big[J'_x S_{i}^x{S}_{i+1}^x
           +J'_y  S_{i}^y {S}_{i+1}^y + J'_z S_{i}^z{S}_{i+1}^z    \nonumber\\
%
%segunda linha
%
& - h' S^{z}_{i} + D'  (S^{z}_{i})^2\big],       \label{1}
\end{align}

\noindent where $S^{\alpha}_i$, $\alpha\in\{x,y,z\}$, 
are the spin matrices at 
the $i$-th site, the $J'_{\alpha}$ are the exchange
interaction couplings  between first neighbors, and
$h'$ is the external magnetic field along the $z$-axis. We also
include the single ion-anisotropy $D'$ parallel to the external
magnetic field.  The periodic boundary condition 
$S^{\alpha}_{N+1}=S^{\alpha}_1$ is used.

In order to render the 
thermodynamic functions to be finite, even in the 
classical limit ($S \rightarrow \infty$),  we define a scaled spin
operator ${\bf s}$   with unitary norm\cite{ejpB}
as ${\bf s} \equiv {\bf S}/\sqrt{S(S+1)}$. 
In order to write explicitly  the effect
of the anisotropy in the $x$ and $y$ directions in relation to the 
$XXZ$ model\cite{ejpB}, we rewrite 
the hamiltonian (\ref{1}) in terms of  the spin operators  $s_i^z$ and  
$s_j^{\pm} \equiv \frac{1}{\sqrt{2}} (s_j^x \pm i s_j^y$), 
with $j=1, \cdots, N$, 

\begin{eqnarray}
{\bf H}& = &      \sum_{i=1}^N 
\{ J[ s_{i}^-  {s}_{i+1}^+ \!   +    s_{i}^+  {s}_{i+1}^-     
+  \! \delta (s_{i}^+  {s}_{i+1}^+   \nonumber \\
%
%segunda linha
%
&& \hspace{-0.8cm}
+ s_{i}^-  {s}_{i+1}^-) 
 +   \Delta s_{i}^z  {s}_{i+1}^z]  
  - hs^{z}_{i} + D (s^{z}_{i})^2\}. \label{2}
\end{eqnarray}

\noindent   The  relations among the  constants in hamiltonians (\ref{1}) 
and (\ref{2}) are: $J \equiv (S/2)(S+1)(J'_x  +  J'_y)$,  
$\delta \equiv  (J'_x -  J'_y)/ (J'_x +  J'_y)$, 
$ \Delta \equiv 2J'_z/ (J'_x +  J'_y)$, 
$h \equiv \sqrt{S (S+1)} h'$ and $D \equiv S(S+1) D'$.

In Ref. \cite{chain_m} we presented  a closed version of the cummulant
expansion for any chain ({any one-dimensional} classical or 
quantum {model}) with  periodic boundary condition and interaction
between first neighbours.   {A survey with the main results  of
\cite{chain_m} is presented in Refs. \cite{PRB03, JPC03}.} In this work we apply that  
method to compute the high temperature
expansion of the HFE  of   the $XYZ$ model
(${\mathcal W}_s$), for a fixed  value of the spin   (with unitary norm). 
Like in the case of the $XXZ$ model,  this  high temperature 
expansion is written as a series  in  powers of $[S(S+1)]^{-1}$.
We use the interpolation method described in Ref. \cite{ejpB}
to calculate the high temperature expansion of ${\mathcal W}_s$,  for 
arbitrary values of $S$, $J$, $\delta$, $\Delta$, $h$ and $D$, up to
order $(J \beta)^5$.  The whole expression is too 
large, so we present it here only up to order $(J \,\beta)^3$,

\begin{widetext}

\begin{eqnarray}
\frac{{\mathcal W}_{s} (\beta)} {J}&=&
-  \frac {\mathrm{ln}(2\,S + 1)}{(J\,\beta) } 
+  \frac {\mathrm{\tilde{D} }}{3}  
+  \left( -  \frac {\delta ^{2}}{9}
 -  \frac {\Delta ^{2}}{18} 
 -  \frac {\tilde{h}^{2}}{6}
 -  \frac {1}{9}
 + ( \frac {1}{30\,S\,(S + 1)}
 - \frac {2}{45})\,\mathrm{\tilde{D}}^{2}  \right)\,(J \,\beta)   \nonumber \\
 &+&  \left( 
 - \frac {\Delta }{36\,S\,(S + 1)} 
 + ( -  \frac {1}{30\,S\,(S + 1)}
 +  \frac {2}{45} )\,\mathrm{\tilde{D}}\,\tilde{h}^{2} 
 + ( \frac {1}{45\,S\,(S + 1)}
  -  \frac {4}{135} )\,\mathrm{\tilde{D}}    \right. \nonumber \\
& + &  \frac {\Delta \,\delta ^{2}}{36\,S\,(S+ 1)}
+  \frac {\Delta \,\tilde{h}^{2}}{9} 
+ ( \frac {1}{126\,S^{2}\,(S + 1)^{2}}
-  \frac {4}{315\,S\,(S + 1)} 
+  \frac {8}{2835} )\,\mathrm{\tilde{D}}^{3}   \nonumber \\
& + &   \left.
( - \frac {1}{45\,S\,(S + 1)} 
+  \frac {4}{135})\,\Delta ^{2}\,\mathrm{\tilde{D}} 
+ ( \frac {1}{45\,S\,(S + 1)}
-  \frac {4}{135} )\,\mathrm{\tilde{D}}\,\delta ^{2}    
               \right) (J\,\beta) ^{2}  \nonumber \\
& + &    \left( 
( -  \frac {1}{5400\,S^{2}\,(S + 1)^{2}}
 +  \frac {2}{675\,S\,(S + 1)} 
-  \frac {7}{2700} )\,\Delta ^{4}     \right.   \nonumber \\
& +  & ( -  \frac {1}{300\,S^{2}\,(S + 1)^{2}} 
 +  \frac {8}{2025\,S\,(S + 1)}
  +  \frac {2}{225} )\,\Delta ^{2}   \nonumber \\
& + &  ( \frac {1}{900\,S^{2}\,(S + 1)^{2}} 
+  \frac {64}{2025\,S\,(S + 1)}  
-   \frac {1}{25} )\,\delta ^{2}   \nonumber \\
& + & ( -  \frac {11}{5400\,S^{2}\,(S + 1)^{2}}  
+  \frac {16}{2025\,S\,(S + 1)} 
-  \frac {1}{1350} )\,\delta ^{4} 
+ ( \frac {1}{60\,S\,(S + 1)} 
+  \frac {2}{135} )\,\tilde{h}^{2}    \nonumber \\
& + & ( -  \frac {1}{84\,S^{2}\,(S + 1)^{2}} 
+  \frac {2}{105\,S\,(S + 1)} 
- \frac {4}{945} )\,\tilde{h}^{2}\,\mathrm{\tilde{D}}^{2}    \nonumber \\
& + & ( -  \frac {3}{700\,S^{2}\,(S + 1)^{2}}
 +  \frac {8}{1575\,S\,(S + 1)}
 +  \frac {4}{4725} )\,\mathrm{\tilde{D}}^{2} 
 + ( -  \frac {1}{540\,S\,(S + 1)} 
 +  \frac {2}{135} )\,\delta ^{2}\,\tilde{h}^{2}   \nonumber \\
& +& ( \frac {1}{90\,S\,(S + 1)}
-  \frac {7}{135} )\,\Delta ^{2}\,\tilde{h}^{2} 
+ ( -  \frac {1}{300\,S^{2}\,(S + 1)^{2}}  
+ \frac {8}{2025\,S\,(S + 1)}  
+  \frac {2}{225} )\,\Delta ^{2}\,\delta ^{2}   \nonumber \\
& + & ( \frac {1}{360\,S\,(S + 1)}  
+  \frac {1}{180} )\,\tilde{h}^{4} 
-  \frac {11}{5400\,S^{2}\,(S + 1)^{2}}  
+  \frac {16}{2025\,S\,(S + 1)} 
 -  \frac {1}{1350}     \nonumber  \\
& + & ( \frac {2}{45\,S\,(S + 1)}  
-  \frac {8}{135} )\,\Delta \,\tilde{h}^{2}\,\mathrm{\tilde{D}}  \nonumber \\
& + & ( \frac {1}{360\,S^{3}\,(S + 1)^{3}}
 -  \frac {97}{18900\,S^{2}\,(S + 1)^{2}}
 +  \frac {8}{4725\,S\,(S + 1)}  
 +  \frac {4}{14175} )\,\mathrm{\tilde{D}}^{4}   \nonumber \\
& + & ( -  \frac {3}{700\,S^{2}\,(S + 1)^{2}} 
 +  \frac {8}{1575\,S\,(S + 1)} 
  +  \frac {4}{4725} )\,\delta ^{2}\,\mathrm{\tilde{D}}^{2}   \nonumber\\
& +  &  \left.
( -  \frac {16}{1575\,S^{2}\,(S + 1)^{2}}  
+  \frac {88}{4725\,S\,(S + 1)} 
- \frac {32}{4725} )\,\Delta ^{2}\,\mathrm{\tilde{D}}^{2}   \right) (J\,\beta) ^{3}
+ \mathrm{O}((J\,\beta )^{4})  ,
\label{3}
\end{eqnarray}

\end{widetext}

\noindent where $\tilde{h} \equiv h/J$ and $\tilde{D} \equiv D/J$.

We point out that these  coefficients of the HFE,  calculated 
up to order $(J\beta)^5$,  are exact and  valid for 
$S = 1/2, 1, 3/2, 2, \cdots$. 
The dependence of the HFE on even 
powers of $\delta$ comes from the
symmetry on the $x$ and $y$  directions.
Letting  ${\mathcal W}_S$  be the  HFE of the $XYZ$ model 
of spin with norm $S(S+1)$, we have that  

\begin{equation}
{\mathcal W}_s (J, \delta, \Delta, h, D; \beta) \equiv 
	{\mathcal W}_S\left(\sigma J, \delta, \Delta, \sqrt{\sigma} h, 
	    \sigma D; \beta\right),
\end{equation}

\noindent where $\sigma=[S(S+1)]^{-1}$.
The HFE is  an homogeneous function of 
first degree, so it is  simple to obtain the
$\beta$-expansion of  ${\mathcal W}_S  (J, \delta, \Delta, h, D; \beta)$ 
from eq. \eqref{3}. For $\delta=0$ we recover the 
results of Ref. \cite{ejpB}.  The full expression of the expansion, 
up to order $(J \, \beta)^5$,  can be obtained 
upon  request to the authors.

It is simple to derive from \eqref{3} the high temperature 
expansion of the specific heat per site of the spin-$S$ 
of the $XYZ$ model, with unitary norm 
($C_s = - \beta^2\frac{\partial^2 (\beta{\mathcal W}_s)}{\partial \beta^2}$).
We obtain $C_s =  -(-\frac{4 D^2}{45} - \frac{h^2}{3}+
\frac{D^2}{15S(S+1)}- \frac{\Delta^2}{9}
- \frac{2}{9}-\frac{2 \delta^2}{9}) \beta^2  + {\cal O} (\beta^3)$,
which shows that in the high temperature region, 
the  $XYZ$ model also 
presents a tail of the Schottky peak\cite{schottky}
($C_{Sch} \propto \beta^2$), for all values of $S$.

\begin{figure}[h!t]
\begin{center}
\includegraphics[width=6cm,height=8.5cm,angle=-90]{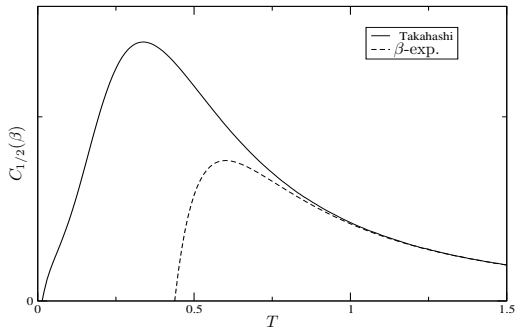}
\caption[fig_1]{  Specif heat of spin-1/2 $XYZ$ chain with 
$h=0$, $D=0$, $J_x= -1.0$,  $J_y= 1.2$ and $J_z = 2.0$. 
The solid line corresponds  to the numerical solution of Takahashi and 
the dashed line represents the $\beta$-expansion of this function
 derived from \eqref{3}.   }
\label{fig_1}
\end{center}
\end{figure}

As a check of our $\beta$-expansion of the HFE \eqref{3}, valid for
arbitrary spin-$S$, we compare the specific heat per 
site for S=1/2  derived from it to the numerical result  obtained
 from the coupled equations by Takahashi\cite{takahashi91}.
 In Fig.\ref{fig_1} we plot the specific heat for $S=1/2$  with $J_x= -1.0$, 
$J_y= 1.2$ and $J_z = 2.0$ (which corresponds to $J = (3/4)\times 0.1$, $\Delta= 20$
and $\delta = -11$) in the absence of an external magnetic field ($h=0$)
and no single-ion anisotropy term ($D=0$). At $T = 0.90$ the difference of the
solution is 2.9\%.

One difficulty  about the spin-$S$ $XYZ$ chain model is that 
it is no longer exactly soluble, in the presence of an external magnetic field,
even for $S=1/2$; moreover, the absence of symmetry in the $x, y$ and $z$
directions makes numerical calculations much more involved.
However, its classical limit, in the high temperature region, 
 can be  easily obtained from eq.\eqref{3} by taking $S\rightarrow \infty$. 

\begin{figure}[h!t]
\begin{center}
 \includegraphics[width=6cm,height=8.5cm,angle=-90]{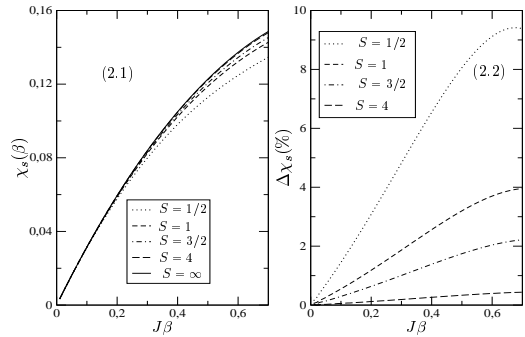}  
\caption[fig_2]{In 2.1 we compare the classical ($S\rightarrow \infty$)
 and the quantum ($S= 1/2, 1, 3/2$ and $4$)
magnetic susceptibility per site as a function of $(J\beta$). In 2.2
we present the relative percental difference between the
quantum and classical results. We let $\Delta= 1$, $\delta= 1$,
$h/J= 0.3$ and $D/J= -0.5$. }
\label{fig_2}
\end{center}
\end{figure}

 In Fig.\ref{fig_2} we show that the magnetic susceptibility 
per site of this model can  be approximated by their classical
behavior even for $S= 3/2$, in the high temperature 
region,  with a percental error larger than 2\%.

\begin{figure}[h!t]
\begin{center}
\includegraphics[width=6cm,height=8.5cm,angle=-90]{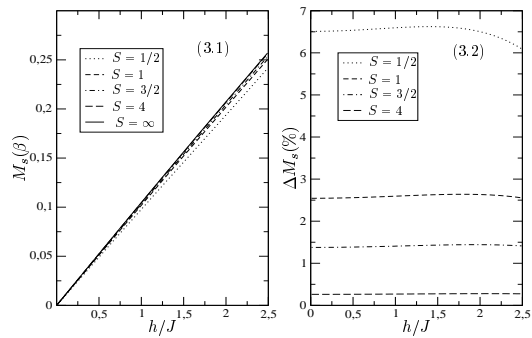}
\caption[fig_3]{
 The magnetization versus $h/J$ at $J\beta = 0.4$ is presented
 in Fig. 3.1 for $S= 1/2, 1, 3/2, 4$ and $S \rightarrow \infty$.
 Figure 3.2 shows the percental difference of those quantum magnetization 
 curves with respect to the classical one. We let $\Delta= 1$, $\delta= 1$, $h/J= 0.3$
  and $D/J= -0.5$.  
 }
\label{fig_3}
\end{center}
\end{figure}

Fig.\ref{fig_3} shows the comparision of the classical magnetization
per site of the model to its quantum version for several values of
spin, in the region of high temperatures. 
We verify that this thermodynamical function 
can be well approximated, in the high  temperature region, 
by its classical result up to $S=1$; the percental
error, in this case, is smaller than 3\% (see Fig. 3.2).

In summary,we have presented the  high temperature expansion
of the HFE of the $XYZ$ model, for arbitrary 
values of spin, in the presence of an external magnetic field and 
single-ion anisotropy term,  up to order $(J\, \beta)^5$. 
The thermodynamic functions derived from eq.\eqref{3}
can be used to fit experimental data to determine
the value of the constants that describe the material under 
interest. As a check, we show that our expansion of the 
specific heat coincides with the  numerical solution of 
Takahashi's coupled equations\cite{takahashi91} up to 
$T\approx 1 $   for $h=0$, $D=0$, $J_x= -1.0$,
$J_y= 1.2$ and $J_z = 2.0$.

We easily obtain the  $\beta$-expansion of the classical
behavior of the model in this region of temperature. 
Finally we showed that the magnetic susceptibility and 
magnetization of the quantum $XYZ$ model can be approximated
by their classical analog for 
$S\geq 3/2 $ and $S\geq 1 $ respectively. 
Our result allows the determination of
the relative percental error  between 
classical and quantum solutions,  for any value of spin, 
 for those thermodynamical functions, although
this is not true for other functions like the specific heat per site.

\begin{acknowledgments}
O. R. thanks FAPEMIG for financial support. The authors are in debt to
CNPq for partial financial support.
S.M. de S. thanks FAPEMIG and M.T.T. thanks FAPERJ for partial
financial support.
\end{acknowledgments}

%=======================================================================

\end{document}